\def\rn{}
\def\nn#1 #2{#2. #1}				
\def\nnn#1 #2 #3{#2. #3. #1}			
\def\nnnn#1 #2 #3 #4{#2. #3. #4 #1}		
\def\nnnnn#1 #2 #3 #4 #5{#2. #3. #4 #5. #1}	
\def\dualand{ and\hbox{ }}				
\def\multiand{, and\hbox{ }}				
\def\rf#1;#2;#3;#4;#5 {{\frenchspacing\par\rn#1, #3 {\bf #4}, #5 (#2). \par}}
\def\rg#1;#2;#3;#4;#5;#6 {{\frenchspacing\par\rn#1, #3 {\bf #4}, #5 (#2). \par}}
\def\rfbook#1;#2;#3;#4;#5 {{\frenchspacing\par\rn#1, {\it #3} (#5, #4, #2).\par}}
\def\rfprep#1;#2;#3 {{\par\frenchspacing\rn#1, #3 (#2).\par}}

\def\beq#1{\begin{equation}\label{#1}}
\def\eeq{\end{equation}}
\def\beqa#1{\begin{eqnarray}\label{#1}}
\def\eeqa{\end{eqnarray}}
\def\eq#1{equation~(\ref{#1})}

\def\spose#1{\hbox to 0pt{#1\hss}}
\def\simlt{\mathrel{\spose{\lower 3pt\hbox{$\mathchar"218$}} \raise 2.0pt\hbox{$\mathchar"13C$}}}
\def\simgt{\mathrel{\spose{\lower 3pt\hbox{$\mathchar"218$}} \raise 2.0pt\hbox{$\mathchar"13E$}}}
\def\simpropto{\mathrel{\spose{\lower 3pt\hbox{$\mathchar"218$}} \raise 2.0pt\hbox{$\propto$}}}

\def\bt{\begin{tabbing}}
\def\et{\end{tabbing}}
\def\beq#1{\begin{equation}\label{#1}}
\def\eeq{\end{equation}}

\def\sec#1{Section~\ref{#1}}

\def\bfig{\begin{figure}[h] \centerline{\hbox{}}\vfill}
\def\efig{\end{figure}\vfill\newpage}

\def\deg{^{\circ}}

\def\l{\ell}
\def\etal{{\frenchspacing\it et al.}}
\def\ie  {{\frenchspacing\it i.e.}}
\def\eg  {{\frenchspacing\it e.g.}}

\documentstyle[prd,aps,epsf,rotate]{revtex}
\begin{document}
\twocolumn[\hsize\textwidth\columnwidth\hsize\csname@twocolumnfalse\endcsname

\title{The Large-Scale Polarization of the Microwave Foreground}

\author{Ang\'elica de Oliveira-Costa$^{1}$, 
                         Max Tegmark$^{1}$, 
                  Christopher O'Dell$^{2}$,
                       Brian Keating$^{3}$,\\
                        Peter Timbie$^{4}$, 
                   George Efstathiou$^{5}$ \&
                        George Smoot$^{6}$}

\address{$^{1}$Department of Physics \& Astronomy, University of Pennsylvania, Philadelphia, PA 19104, USA, 
	       angelica@higgs.hep.upenn.edu\\}
\address{$^{2}$Department of Astronomy, University of Massachusetts, Amherst, MA 01003, USA\\}
\address{$^{3}$Department of Physics, California Institute of Technology, Pasadena, CA 91125, USA\\}
\address{$^{4}$Department of Physics, University of Wisconsin, Madison, WI 53706-1390, USA\\}
\address{$^{5}$Institute  of Astronomy, University of Cambridge, Cambridge CB3 OHA, UK\\}
\address{$^{6}$Department of Physics, University of California, Berkeley, CA 94720, USA\\}


\maketitle


\begin{abstract}
Most of the useful information about inflationary gravitational 
waves and reionization is on large angular scales where Galactic 
foreground contamination is the worst, so a key challenge is to model, 
quantify and remove polarized foregrounds.
We use the Leiden radio surveys to quantify the polarized synchrotron 
radiation at large angular scales, which is likely to be the most 
challenging polarized contaminant for the WMAP satellite.
We find that the synchrotron $E$- and $B$-contributions are equal to 
within 10\% from $408-820$MHz with a hint of $E$-domination at higher 
frequencies. We quantify Faraday Rotation \& Depolarization effects 
and show that they cause the synchrotron polarization percentage 
to drop both towards lower frequencies and towards lower multipoles.
\bigskip
\end{abstract}

]
  

\section{INTRODUCTION}

CMB polarization and its decomposition into $E$ and $B$ modes is a topic
of growing importance and interest in cosmology\cite{zalda03}. In the era 
of WMAP \cite{BennettMission}, a key issue is to estimate the contribution of 
Galactic foregrounds (more specifically, polarized synchrotron emission) 
at the large angular scales.
Unfortunately, these large scales are also the ones where polarized 
foreground contamination is likely to be most severe, both because of the 
red power spectra of diffuse Galactic synchrotron and dust emission and 
because they require using a large fraction of the sky, including less 
clean patches. The key challenge in the CMB polarization endeavor will 
therefore be modeling, quantifying and removing large-scale polarized 
Galactic foregrounds.
 
Unfortunately, we still know basically nothing about the polarized 
contribution of the Galactic synchrotron component at CMB frequencies
\cite{foregpars,tucci00,Baccigalupi00,Burigana02,bruscoli02,tucci02,giardino02}, 
since it has only been measured at lower frequencies and extrapolation 
is complicated by Faraday Rotation. This is in stark contrast to the CMB 
itself, where the expected polarized power spectra and their dependence on 
cosmological parameters has been computed from first principles to high 
accuracy \cite{K97,ZS97,Z98,HuWhite97}.

This is the topic of the present proceeding. We will employ polarization 
sensitive radio surveys to further quantify the polarized synchrotron 
radiation, which is likely to be the most challenging contaminant 
in the polarization maps expected from the WMAP satellite 
\cite{BennettForegs,kogut03}.
This proceeding is organized as follows: in section \sec{synchro}, 
we review the basics of the synchrotron emission, as well the problems
involved with extrapolations from lower to higher frequencies. We 
present our results as well as discuss our conclusions in \sec{ResultsSec}. 
For more details about this analysis consulte \cite{doc02}


\section{Our Knowledge of Synchrotron Emission}\label{synchro}

The Galactic InterStellar Medium (ISM) is a highly complex medium 
with many different constituents interacting through a multitude of 
physical processes. Free electrons spiraling around the Galactic 
magnetic field lines emit synchrotron radiation \cite{rybicki}, 
which can be up to 70\% linearly polarized (see \cite{davies98,smoot99} 
for a review). 

The power spectrum $C_\l$ of synchrotron radiation is normally modeled 
as a power law in both multipole $\l$ and frequency $\nu$, which we will 
parametrize as 
\beq{freq_dep}
    \delta T_\l^2(\nu) = 
    A \left( {\l\over 50} \right) ^{\beta+2} {\rm with}\quad 
    A\propto \nu^{2\alpha},
\eeq
where $\delta T_\l\equiv [\l(\l+1)C_\l/2\pi]^{1/2}$. This definition implies 
that $C_\l\propto\l^\beta$ for $\l\gg 1$ and that the fluctuation amplitude 
$\propto\nu^{\alpha}$. The standard assumption is that the unpolarized 
intensity has $\alpha\approx -2.8$ with variations of order $0.15$ across 
the sky \cite{platania} -- see also \cite{banday91,jonas99,roger99}. 

As to the power spectrum slope $\beta$, the 408~MHz Haslam map 
\cite{haslam1,haslam} suggests $\beta$ of order -2.5 to -3.0 down to 
its resolution limit of $\sim 1^\circ$ \cite{TE96,bouchet96,bouchet99,newspin}. 
A similar analysis done on the 2.3~GHz Rhodes map of resolution 
20$^\prime$ \cite{jonas99} gives $\beta = -2.92\pm 0.07$ \cite{giardino01} 
(flattening to $\beta\approx -2.4$ at low Galactic latitudes \cite{giardino02}). 

For the polarized synchrotron component, our observational knowledge is, 
unfortunately, not as complete.  To date, there are measurements of the 
polarized synchrotron power spectrum obtained basically from three 
different surveys \cite{reich01}: 
	the Leiden surveys \cite{BS76,S84}, 
         the Parkes 2.4~GHz Survey of the Southern Galactic Plane \cite{D95,D97}, 
         and the Medium Galactic Latitude Survey \cite{U98,U99,D99}.
These measurements exhibit a much bluer power spectrum in polarization 
than in intensity, with $\beta$ in the range from 1.4 to 1.8 
\cite{foregpars,tucci00,Baccigalupi00,Burigana02,bruscoli02,tucci02,giardino02}. 
These results are usually taken with a grain of salt when it comes to their 
implications for CMB foreground contamination, for three reasons: 

\medskip 
\noindent{\it{1.}} Extrapolations are done from low to high latitudes;\\ 
	 {\it{2.}} Extrapolations are done from low to high frequencies;\\
	 {\it{3.}} Much of the available data is undersampled.  
\medskip 

The Leiden surveys extend to high Galactic latitudes and up to 1.4 GHz 
but are unfortunately undersampled, while the Parkes and the Medium 
Galactic Latitude Surveys only probe regions around the Galactic 
plane -- see \cite{doc02} for more details. In the following three 
subsections, we will discuss these three problems in turn.

\subsection{The Latitude Extrapolation Problem}

There is a well-know empirical result that shows that whereas the unpolarized 
synchrotron emission (at MHz range) depends strongly on the Galactic latitude, 
the polarized component is approximately independent of Galactic latitude 
(see, \eg, \cite{D97}). 
The usual interpretation for this very weak latitude dependence of polarized
synchrotron radiation is that the signal is dominated by sources that are nearby
compared to the scale height of the Galactic disk, with more distant sources 
being washed out by Depolarization (to which we return in the next subsection).
As a result, having well-sampled polarized maps off the galactic plane at the 
same frequencies would not be expected to affect our results much, since they 
would be similar to those in the plane. 
This issue, however, deserves more work as far as extrapolation to CMB 
frequencies is concerned: the latitude dependence may well return at higher 
frequencies as Depolarization becomes less important,  thereby revealing 
structure from more distant parts of the Galactic plane. In this case, 
extrapolating from an observing region around the Galactic plane to higher 
latitudes may well result in less small-scale power in the angular 
distribution.
 
\setcounter{footnote}{1}

\subsection{ Frequency Extrapolation Problem}
\label{faraday}

It is important to point out, that Faraday Rotation (see, \eg, 
\cite{sokoloff98}) can only change the polarization angle and 
not the polarized intensity $P$ ($P$=$\sqrt{Q^2+U^2}$). The 
fact that we do see structure in $P$ that is not correlated 
with a counterpart in intensity $T$ implies that part of the 
radiation has been depolarized \cite{wieringa93}.  
Depending on the frequency and beamwidth used, Depolarization can
play an important role in polarization studies of the Galactic radio 
emission \cite{S84} -- for more details, see Cortiglioni and Spoelstra
\cite{CS95}.
Because of the complicated interplay of these mechanisms, we should 
expect both the amplitude and the shape of the polarized synchrotron 
power spectrum to change with frequency. 

\bigskip
\bigskip
\bigskip
\bigskip
{\bf \centerline{Table 1 -- Normalization \& Spectral Index$^{(a)}$}} 
\medskip
\centerline{
\begin{tabular}{ccccc}
\hline
\hline
\multicolumn{1}{c}{$\nu$}       &
\multicolumn{1}{c}{$A_E$}       &
\multicolumn{1}{c}{$\beta_E$}   &
\multicolumn{1}{c}{$A_B$}       &
\multicolumn{1}{c}{$\beta_B$}   \\
 (GHz)    &$[mK^2]$ &	     &$[mK^2]$ &    \\
\hline
 0.408    &5.5      &-0.5    &5.7      &-0.4\\
 0.465    &5.4      &-1.0    &5.4      &-0.5\\
 0.610    &5.1      &-1.0    &5.1      &-0.8\\
 0.820    &4.5      &-1.5    &4.6      &-1.8\\
 1.411    &3.9      &-1.9    &3.6      &-2.6\\
\hline
\hline
\end{tabular} 
}
\medskip
\noindent{\small $^{(a)}$All fits are normalized at $\ell$=50, \ie,
		        $\delta {\rm T}_{\ell}^2 = A (\ell/50)^{\beta+2}$.} 
\smallskip

\subsection{Incomplete Sky Coverage and the Undersampling Problem}

For the case of undersampling in the Leiden surveys, some authors have
overcome this problem by doing their Fourier analysis over selected
patches in the sky where they believe the average grid space in the 
patch is close to the map's beam size, so that they can apply a 
Gaussian smoothing on it -- this is well explained and illustrated 
in \cite{bruscoli02}.
Fortunately, we can eliminate this problem by measuring the power 
spectra with the matrix-based quadratic estimator technique that has 
recently been developed for analyzing CMB maps \cite{BJK,TC01,angel_pique}.
Although the undersampling and partial sky coverage results in 
unavoidable mixing between different angular scales $\l$ and 
polarization types ($E$ and $B$), this mixing (a.k.a. {\it leakage}) 
is fully quantified by the window functions that our method computes 
\cite{TC01} and can therefore be included in the statistical analysis 
without approximations. Specifically, we compute the six power spectra 
	($C_\l^T    \equiv T,
	  C_\l^E    \equiv E,
	  C_\l^B    \equiv B,
	  C_\l^{TE} \equiv X,
	  C_\l^{TB} \equiv Y,
	  C_\l^{EB} \equiv Z$) 
so that the much discussed \cite{TC01,Z98,Jaffe00,Zalda01,Lewis02,Bunn01,Bunn02}  
$E-B$ leakage is minimal \cite{angel_pique}.


\section{Results \& Conclusions}\label{ResultsSec}

We employed only the Leiden surveys \cite{BS76,S84} for our analysis. The observations 
done by Brouw and Spoelstra covered almost 40\% of the sky extending to high 
Galactic latitudes. Using the same instrument, they observed the polarized Galaxy in 
 $Q$ and $U$ in five frequencies from 408~MHz up to 1.4~GHz and with angular 
resolutions from 2.3$\deg$ at 408~MHz up to 0.6$\deg$ at 1.4GHz. Unfortunately 
this data was also undersampled, making it difficult to draw inferences about its 
polarized power spectrum.
               
Using matrix-based quadratic estimator methods \cite{TC01,angel_pique}, we measure 
the power spectra from the Leiden surveys, obtaining the following key results:
\begin{enumerate}
	\item Our analysis was performed using 10 multipole bands of width $\Delta
	      \l=10$ for each of the six polarization types $(T,E,B,X,Y,Z)$, thereby 
	      going out to $\l=100$. We used the Haslam map for the unpolarized 
	      component $T$, scaled and smoothed to match Leiden's five different 
	      frequencies, and assuming a  $|b|=25\deg$ Galactic cut.
	      The best fit normalizations $A$ and slopes $\beta$ for $E$ and $B$ are 
	      shown in Table~1. The values of $\beta$ are consistent with previous analyses 
	      \cite{foregpars,tucci00,Baccigalupi00,Burigana02,bruscoli02,tucci02,giardino02}, 
	      showing that the slopes get redder as frequency increases.
	      For all Leiden surveys, the $X$ and $Y$ power spectra are found to be 
	      consistent with zero -- the 2.4~GHz Parkes survey had a similar finding 
	      for $X$ \cite{giardino02}. These are not surprising results: if Faraday 
	      Rotation makes the polarized and unpolarized components to be uncorrelated, 
	      it is natural to expect that $X,Y$=0. However, at the CMB frequencies 
	      (where the effects of Faraday Rotation \&  Depolarization are unimportant) 
	      this should not be the case.
	\item To study the frequency dependence, we average the 10 multipole bands 
	      of the Leiden power spectrum measurements together into a single band 
	      for each polarization type to reduce noise. From these results, we 
	      fit the average frequency dependence (for the $25\deg$ cut data) as a 
	      power law as in \eq{freq_dep} with slope $\alpha_E = -1.3$ and 
	      $\alpha_B = -1.5$ for $E-$ and $B-$polarization, respectively.
	\item An interesting question about polarized foregrounds is how their 
	      fluctuations separate into $E$ and $B$. Although many authors 
	      initially assumed that foregrounds would naturally produce equal 
	      amounts of $E$ and $B$, Zaldarriaga \cite{Zalda01} showed that 
	      this need not be the case. 
	      Early studies \cite{Baccigalupi00,giardino02} have indicated 
	      that $E\approx B$ at 2.4~GHz in the Galactic plane. However, 
	      these analyses used Fourier transforms and spin-2 angular 
	      harmonic expansions, respectively, without explicitly computing 
	      the window functions quantifying the leakage between $E$ and $B$.
	      We therefore perform a likelihood analysis of the Leiden surveys 
	      specifically focusing on this question, and including an exact 
	      treatment of the leakage. The likelihood analysis of the data 
	      is done with two free parameters corresponding to the overall 
	      normalization of the $E$ and $B$ power spectra, and assuming 
	      that they both have the same power law shape given by the slopes 
	      $\beta_E$ from Table~1.
	      We obtain that the synchrotron $E$- and $B$-contributions are 
	      equal to within 10\% from 408 to 820~MHz, with a hint of 
	      $E$-domination at higher frequencies. One interpretation is 
	      that $E>B$ at CMB frequencies but that Faraday Rotation mixes 
	      the two at low frequencies.
	\item Faraday Rotation \& Depolarization effects depend not only on 
	      frequency but also on angular scale -- they are important at 
	      low frequencies ($\nu\simlt 10$ GHz) and on large angular scales.
	      Therefore, we must take into account Faraday Rotation \& Depolarization 
	      effects when extrapolating radio survey results from low to 
	      high galactic latitudes and from low to high frequencies.
	\item We detect no significant synchrotron $X$ cross 
	      correlation, but Faraday Rotation could have hidden a substantial 
	      correlation detectable at CMB frequencies.  
	\item Combining the POLAR \cite{keating02,odell02,doc02} and radio 
	      frequency results, and the fact that the $E$-polarization of 
	      the abundant Haslam signal in the POLAR region is not detected 
	      at 30 GHz, suggests that the synchrotron polarization percentage 
	      $p$ at CMB frequencies is rather low ($p<$20\%).
\end{enumerate}
Experiments such as polarized WMAP and Planck will shed significant new light 
on synchrotron polarization and allow better quantification of its impact both on 
these experiments and on ground-based CMB observations.


\bigskip
\bigskip

This work was supported by 
NSF grants AST-0071213 \& AST-0134999 and
NASA grants NAG5-9194 \& NAG5-11099.
MT acknowledges a David and Lucile Packard Foundation fellowship
and a Cottrell Scholarship from Research Corporation.



\end{document}